\documentclass[pra,twocolumn, superscriptaddress]{revtex4-1}
\usepackage{graphicx,amsmath,amssymb,amsxtra,epstopdf,wasysym}
\usepackage{mathrsfs}
\usepackage{hyperref}   
\usepackage{float}

\renewcommand{\narrowtext}{\begin{multicols}{2} \global\columnwidth20.5pc}

\def\be{\begin{eqnarray}}
\def\ee{\end{eqnarray}}

\def\RVB19{3280}
\begin{document}

\title{Floquet edge states in a harmonically driven integer quantum Hall system}

\author{Zhenyu Zhou}
\affiliation{
Department of Physics and Astronomy,
University of Pittsburgh, Pittsburgh, PA 15260}
\affiliation{
School of Physics, Astronomy and Computational Sciences, George Mason University, Fairfax, VA 22030}

\author{Indubala I. Satija}
\affiliation{
School of Physics, Astronomy and Computational Sciences, George Mason University, Fairfax, VA 22030}

\author{Erhai Zhao}
\affiliation{
School of Physics, Astronomy and Computational Sciences, George Mason University, Fairfax, VA 22030}

\begin{abstract}
Recent theoretical work on time-periodically kicked Hofstadter model found robust counter-propagating edge modes. 
It remains unclear how ubiquitously such anomalous modes can appear, and what dictates their robustness against disorder. Here we shed further light on the nature of these modes by analyzing a simple type of periodic driving where the hopping along one spatial direction is modulated sinusoidally with time while the hopping along the other spatial direction is kept constant. We obtain the phase diagram for the quasienergy spectrum at flux 1/3 as the driving frequency $\omega$ and the hopping anisotropy are varied. A series of topologically distinct phases with counter-propagating edge modes appear due to the harmonic driving, similar to the case of a periodically kicked system studied earlier. We analyze the time dependence of the pair of Floquet edge states localized at the same edge, and compare their Fourier components in the frequency domain. In the limit of small modulation, one of the Floquet edge mode within the pair can be viewed as the edge mode originally living in the other energy gap shifted in quasienergy by $\hbar \omega$, i.e., by absorption or emission of a ``photon" of frequency $\omega$. Our result suggests that counter-propagating Floquet edge modes are generic features of periodically driven integer quantum Hall systems, and not tied to any particular driving protocol. It also suggests that the Floquet edge modes would remain robust to any static perturbations that do not destroy the chiral edge modes of static quantum Hall states.

\end{abstract}

\maketitle
\section{Introduction}
The edge states of a stationary topological insulator (or topological superconductor) are intimately related to its bulk topological properties \cite{RevModPhys.82.3045,RevModPhys.83.1057}. This well established relationship is known as the bulk-boundary correspondence. The bulk topological invariants by definition are insensitive to small perturbations as long as the gaps remain open and the underlying discrete symmetries are preserved \cite{schnyder_classification_2008,kitaev_periodic_2009,PhysRevB.85.085103}. Accordingly, the edge states are robust against such perturbations and protected by the corresponding symmetries. In this paper, we explore robust edge states in time-periodic quantum systems, for which the bulk-boundary correspondence relationships are much less understood. We focus on a simple model of non-interacting, spinless fermions on the square lattice, the Hofstadter model \cite{hofstadter_energy_1976} with time-periodically modulated hopping between nearest neighbors. This model could potentially be realized in cold atoms experiments with synthetic gauge field \cite{lin_synthetic_2009,dalibard_colloquium:_2011} and periodically modulated optical lattices \cite{eckardt_superfluid-insulator_2005,hemmerich_effective_2010}. Here we use it primarily as a toy model to investigate two-dimensional topological phases of matter under periodic driving.

Several important theoretical results have been established recently regarding the topological properties of periodically driven systems. 
The early work by Oka and Aoki \cite{oka_photovoltaic_2009}, Lindner et al. \cite{lindner_floquet_2011}, and Kitagawa et al. \cite{kitagawa_topological_2010} have explicitly shown how periodical driving can turn topologically trivial band insulators into Floquet topological insulators. Jiang et al. \cite{jiang_majorana_2011} found Floquet Majorana fermions in periodically driven one-dimensional cold atoms with $s$-wave attractive interaction. Many following work explored Floquet topological insulators and superconductors in a wide variety of driven systems \cite{kitagawa_transport_2011,gu_floquet_2011,dora_optically_2012,ho_quantized_2012,suarez_morell_radiation_2012,liu_floquet_2013,lindner_topological_2013,rudner_anomalous_2013,kundu_transport_2013,katan_modulated_2013,thakurathi_floquet_2013,delplace_merging_2013,lababidi_counter-propagating_2014,wang_floquet_2014,reichl_floquet_2014,zhou_aspects_2014,ho_effects_2014}. Experimentally, Floquet edge states have been demonstrated in photonic crystals \cite{rechtsman_photonic_2013} and photonic quantum walks \cite{kitagawa_observation_2012}. In particular, periodically modulated (shaken) optical lattice has been used experimentally to engineer the band structure \cite{parker_direct_2013,zheng_floquet_2014,zhang_shaping_2014} and generate artificially gauge fields \cite{struck_engineering_2013,hauke_non-abelian_2012,baur_dynamic_2014} for ultracold atoms. 
In the limit of fast driving, a periodically driven system can often be described by an effective Hamiltonian, $\mathscr{H}_{eff}$, which can be subsequently analyzed in the same way as any stationary Hamiltonians  \cite{lindner_floquet_2011,kitagawa_topological_2010}. For example, they can be classified and understood according to the spatial dimension and the underlying discrete symmetries in the framework of symmetry protected topological phases \cite{schnyder_classification_2008,kitaev_periodic_2009,PhysRevB.85.085103}. A key insight recognized in recent theoretical work is that the topological properties of periodically driven systems can be unique and have no stationary analogues. In general, they cannot be fully described by the effective Hamiltonian. Kitagawa et al. \cite{kitagawa_topological_2010} have shown that the Floquet operator $\mathscr{U}(T)$, defined as the time evolution operator $\mathscr{U}$ over the full period $T$, can be used to construct the topological invariants for driven lattice fermions in one and three dimensions. Rudner et al. \cite{rudner_anomalous_2013} constructed the topological invariants, the winding numbers, in two dimensions (2D) based on the integral of time evolution operator $\mathscr{U}(0\leq t\leq T)$. 

It is worthwhile to stress the difference between the spectra of static systems and periodically driven systems. The energy spectra of stationary band insulators are typically bounded from above and below, $E\in (E_L, E_H)$, for lattice systems. Edge states are confined in the band gaps in between. For example, a system with $q$ bands has $q-1$ band gaps. In contrast, the quasienergy spectrum of periodically driven system lives in the quasienergy Brillouin zone (QBZ), $\epsilon\in [-\pi/T, \pi/T]$, which is topologically a closed circle. This has an immediate consequence. A system with $q$ quasienergy bands has $q$ band gaps. Thus, it can support {\it more edge states} compared to an analogous stationary system. For example, imagine there is a gap at the quasienergy Brillouin zone boundary, $\epsilon\sim \pm \pi/T$. Then it is possible to have the so-called ``$\pi$-modes" inside such  ``$\pi$-gap" ($\pi$ here refers to the fact that $\epsilon T\sim \pm \pi$, i.e., the mode lives near the QBZ boundary). For driven 2D systems, each of the quasienergy gap is characterized by a winding number $w$ describing the topological property of $\mathscr{U}(t)$ \cite{rudner_anomalous_2013}. As an example of bulk-boundary correspondence of driven systems, $w$ equals to the net chirality, i.e., the number of right moving edge modes minus the number of left moving edge modes within the same quasienergy gap \cite{rudner_anomalous_2013}. Note that no additional symmetry of $\mathscr{U}$ was assumed in deriving this important result.

Most previous work on driven 2D systems, e.g., Refs. \onlinecite{oka_photovoltaic_2009,lindner_floquet_2011,kitagawa_topological_2010,kitagawa_transport_2011,gu_floquet_2011, katan_modulated_2013}, focused on how driving turns topologically trivial band insulators into topological insulators. To further explore the edge states and bulk-boundary correspondence in periodically driven 2D systems, we have investigated periodically kicked Hofstadter model at fixed flux $e/3h$ per plaquette \cite{lababidi_counter-propagating_2014}, which in the static limit is arguably the simplest topological insulator in 2D. Most interestingly, we found that despite the winding number being zero inside the $\pi$-gap, there are robust counter-propagating edge modes  \cite{lababidi_counter-propagating_2014}. Static perturbations such as random onsite potential cannot gap out these $\pi$-modes. In particular, there exist phases with the same set of winding and Chern numbers but distinctive edge state spectra  \cite{lababidi_counter-propagating_2014}. It suggests that other topological invariants, in additional to the winding numbers, may be constructed to predict the existence of these modes and describe their stability. These results have been independently confirmed in related kicked Harper model \cite{ho_effects_2014} and 
continuously driven Harper models \cite{zhou_aspects_2014}.

The purpose of this paper is to test the hypothesis that counter-propagating edge mode is a generic feature of driven integer quantum Hall system and not tied to a particular class of driving protocols. We also shed more light on the nature of the $\pi$-modes by relating them to the well known chiral edge mode of static integer quantum Hall insulators using a perturbative construction. We achieve these two goals by analyzing the Hofstadter model under harmonic (i.e., sinusoidal) driving using the standard Floquet analysis in the frequency domain.

\section{Harmonically driven Hofstadter model} 

We consider a tight-binding model of single spin species fermions hopping on a square lattice in the presence of a fixed, homogeneous magnetic field. The hopping amplitudes along the $x$ and $y$ direction, $J_x$ and $J_y$, are periodic functions of time $t$ with period $T$. The Hamiltonian has the form
\begin{equation}
\label{Ht}
H(t)=-J_x(t)\sum_{\mathbf{r}} c_{\mathbf{r}+\hat{x}}^{\dagger}c_\mathbf{r}-J_y(t)\sum_r e^{i2\pi x \alpha} c_{\mathbf{r}+\hat{y}}^{\dagger}c_\mathbf{r}+ h.c. 
\end{equation}
Here the position vector $\mathbf{r}$ labels the lattice site, $\mathbf{r}=x\hat{x}+y\hat{y}$, and $c_\mathbf{r}^{\dagger}$ is the creation operator for site $\mathbf{r}$ (we take the lattice constant as our length unit). We work in the Landau gauge, and $\alpha$ is the magnetic flux per plaquette measured in units of flux quantum $h/e$. Model Eq.  \eqref{Ht} reduces to the well known Hofstadter model when both $J_x$ and $J_y$ are constants \cite{hofstadter_energy_1976}. Lattice models with time-periodically modulated hoppings have been studied, e.g., in Ref. \onlinecite{kitagawa_topological_2010} and Ref. \onlinecite{rudner_anomalous_2013} in the context of Floquet topological insulators.

Previously, we have studied model Eq. \eqref{Ht} in great detail for one particular type of driving protocols in which $J_{x,y}(t)$ take the form of alternating square waves  \cite{lababidi_counter-propagating_2014}. The periodic driving is characterized by two independent driving parameters $\theta_x$ and $\theta_y$, which are nothing but the time integrals of $J_x(t)$ and $J_y(t)$ over the whole period $t\in [0,T]$ (we set $\hbar=1$ throughout this paper). In the limit of $J_x$ being constant and $J_y$ becoming a Dirac comb function, $J_y(t)\propto \sum_n \delta(t-nT)$, the system is periodically kicked, just as in a periodically kicked rotor. For simplicity, we shall refer to the type of model consider in Ref.  \onlinecite{lababidi_counter-propagating_2014} as {\it periodically kicked} Hofstadter model. 

In this paper, we study another type of time modulation where $J_x$ stays constant while $J_y$ varies sinusoidally with time,
 \begin{equation}
 \label{mod}
 J_x(t)=j_x, \;\;\;\; J_y(t)=j_y\left[1+\gamma \cos(\omega t)\right].
 \end{equation}
Here $\omega$ is the driving frequency, and the period $T=2\pi/\omega$. The dimensionless parameter $\gamma$ controls the strength of modulation. For example, $\gamma=0$ is the limit of the static Hofstadter model, while strong modulation such as $\gamma=1$ resembles periodic kicking. To facilitate the comparison, we shall refer to this type of model considered here as {\it harmonically driven} Hofstadter model.

Our motivation to study the harmonically driven Hofstadter model is two-fold. First of all, we would like to confirm that  counter-propagating edge modes and the various topologically distinct phases found for the periodically kicked model also arise in harmonically driven integer quantum Hall systems. Secondly and more importantly, as we show below using Floquet analysis in the frequency domain, the quasienergy spectra including the edge states are easier to visualize and yield additional insights. It is smoothly connected to the static model by taking $\gamma\rightarrow 0$. In contrast, periodically kicked systems are easier to analyze directly in the time domain, for example, by constructing the Floquet operator $\mathscr{U}(T)$. The analysis in this paper is thus complementary to that in Ref. \onlinecite{rudner_anomalous_2013}.

Now we outline how model Eq. \eqref{Ht}  can be solved  by Floquet analysis \cite{PhysRev.138.B979,PhysRevA.7.2203}. By definition, $H$ is periodic in time,
\[
H(t)=H(t+T).
\]
The Floquet theorem states that the wave function has the form $\psi_{n}(t)=e^{-i\epsilon_n t}\phi_n(t)$, where $\epsilon_n$ is the quasienergy and $\phi_n(t+T)=\phi_n(t)$. This allows the expansion of $\phi_n(t)$ into Fourier series in terms of fundamental frequency $\omega$,
\begin{equation}
\label{FT}
\psi_{n}(t)=e^{-i\epsilon_n t}\sum_{m=-\infty}^{\infty}\phi_{n,m}e^{im\omega t}.
\end{equation}
Substituting Eq. \eqref{FT} into the time dependent Schrodinger equation $i\partial_t \psi (t)=H(t)\psi(t)$, we find that it is reduced to
a time-independent eigenvalue problem \cite{PhysRevA.7.2203},
\begin{equation}
\label{HF}
\sum_{m'}\mathcal{H}_{m,m'}\phi_{n,m'}=\epsilon_n\phi_{n,m},
\end{equation}
where the Floquet Hamiltonian
\begin{equation}
\mathcal{H}_{m,m'}=m\omega \delta_{mm'}+\frac{1}{T}\int_0^T dt \, e^{-i(m-m')\omega t}H(t).
\end{equation}
$\mathcal{H}$ lives in a Hilbert space \cite{PhysRevA.7.2203} spanned by the complete basis $\{|\beta\rangle\otimes|m\rangle\}$, where $|\beta\rangle$ is some basis for the static Hofstadter model (e.g., $|\mathbf{r}\rangle$ or $|\mathbf{k}\rangle$) and $|m\rangle$ takes the form of $e^{im\omega t}$ in the time domain. We call $\phi_{n,m}$ the $m$-th Floquet mode corresponding to quasienergy $\epsilon_n$.  
For harmonic driving, Eq. \eqref{mod}, it is easy to see that $\mathcal{H}$ takes a tri-block-diagonal form,
\[
\mathcal{H}_{m,m'}=[m\omega +H_0]\delta_{mm'}+\gamma H_1 (\delta_{m,m'+1}+\delta_{m+1,m'}).
\]
We shall refer to $m\omega +H_0$ as the $m$-th subband, since it is nothing but the static Hofstadter model, $H_0$, shifted in energy by $m\omega$.
The driving term proportional to $\gamma$ only couples neighboring subbands, i.e., $m$ to $m\pm 1$.  The Floquet Hamiltonian is formally very much analogous to that of an electronic crystal in the presence of monochromatic light field where photons drive vertical transitions between neighboring bands, leading to band mixing. For this reason, we can then describe the effect of harmonic driving, in the spirit of perturbation theory, in terms of fermions in a given subband emitting or absorbing a ``photon" of frequency $\omega$. We note that for periodically kicked systems with square wave or delta function driving  \cite{lababidi_counter-propagating_2014}, the $m$-th subband is coupled to infinitely many other subbands.

To solve the eigenvalue problem Eq. \eqref{HF}, we follow the standard practice to truncate the Floquet Hamiltonian $\mathcal{H}_{m,m'}$ into a finite dimensional matrix with $-M\leq m,m' \leq M$ \cite{rudner_anomalous_2013}. This is justified because the Floquet state $\phi_{n,m}$ decays rapidly with $|m|$ beyond a finite range in frequency space. For example, $M=5$ already yields accurate results. We always check that $M$ is chosen to be large enough so that the results do not depend on the truncation as $M$ is further increased. 

\begin{figure}[t]
\centering
\includegraphics[width=0.48\textwidth]{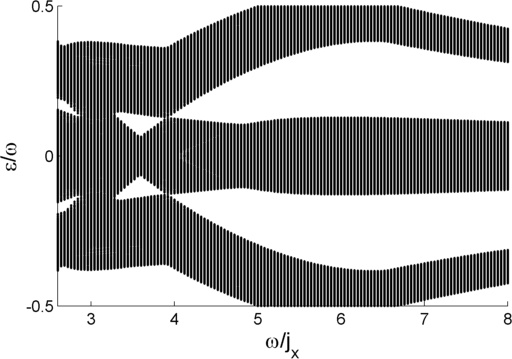}
\caption{
\label{bulk-gap}
The bulk quasienergy spectrum $\epsilon$ versus the driving frequency $\omega$ for $j_y/j_x=1.5$. We focus on intervals of $\omega$ for which there are three well defined gaps in the quasienergy Brillouin zone $\epsilon \in [-\omega/2, \omega/2]$. 
}
\end{figure}

\begin{figure}[h!]
\centering
\includegraphics[width=0.48\textwidth]{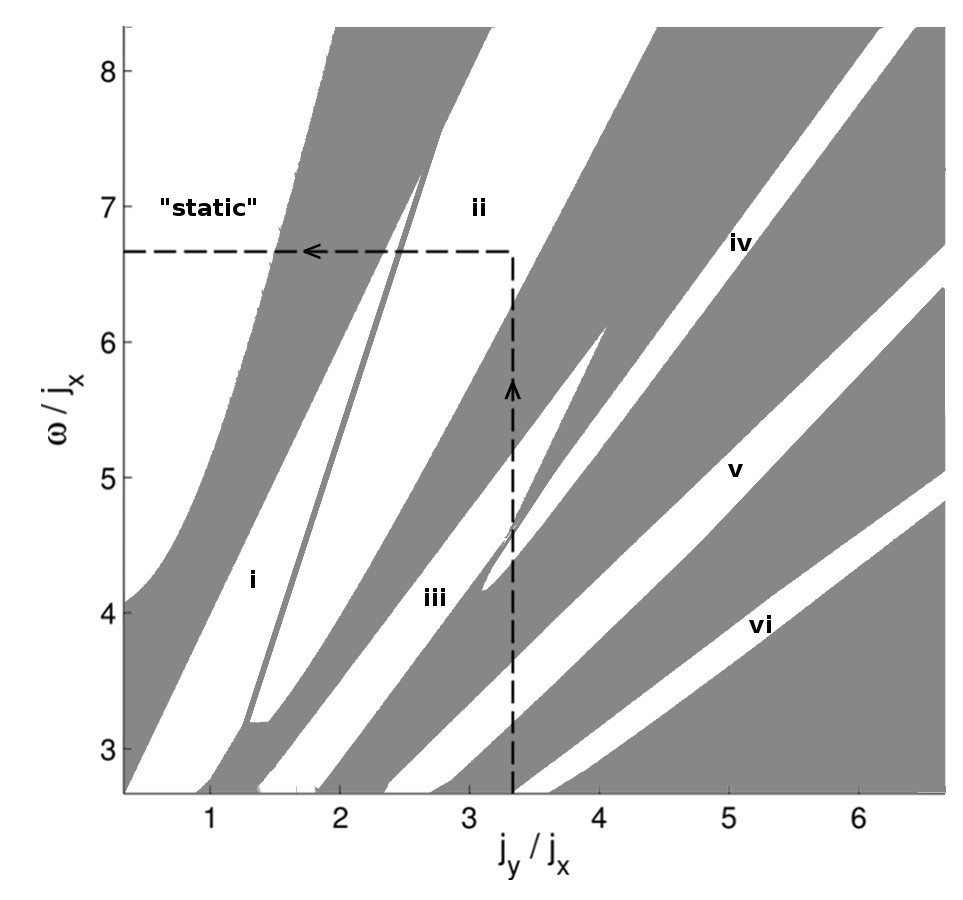}
\caption{
\label{2dpd}
The phase diagram of harmonically driven Hofstadter model, Eqs. \eqref{Ht} and \eqref{mod}, in the $(j_y, \omega)$ plane for $\alpha=1/3$, and $\gamma=1$. The white regions are phases with three well defined quasienergy bands and three gaps. The dash line depicts a cut through the phase diagram. The representative spectra of phase i to vi are shown in Fig. \ref{samples}. Each phase has its characteristic Chern numbers and edge state spectrum. They are summarized in Fig. \ref{1dpd}.  }
\end{figure}

\section{Phase diagram and the quasienergy spectrum}

We focus on the harmonically driven Hofstadter model for the simple case of $\alpha=1/3$, so direct comparison with Ref. \onlinecite{lababidi_counter-propagating_2014} can be easily established. The energy spectrum for the static Hofstadter model at flux 1/3 consists of three bands (e.g., see top panel of Fig. 4) since the magnetic unit cell contains three lattice sites. A quantum Hall insulator is realized when the Fermi energy lies within one of the band gaps. For each physical edge of the sample, exactly one chiral edge mode transverses the energy gap. 
The quasienergy spectrum of the harmonically driven Hofstadter model are much more complicated. Examples of the bulk spectrum are shown in Fig. \ref{bulk-gap} for different driving frequencies.
We focus on ``phases" that have three well defined quasienergy bands (and thus three gaps). Note that it is sufficient to present the quasienergy spectrum in the quasienergy Brillouin zones (QBZ),  $\epsilon \in [-\omega/2,\omega/2]$. The three
bands within the QBZ, from top to bottom, have Chern number $c_1$, $c_2$, and $c_3$ respectively. Accordingly, we refer to the three gaps from the top to the bottom of the QBZ as gap1 (the $\pi$-gap), gap2, and gap3. The total number of edge modes at each physical edge, regardless their group velocity, within gap1 and gap2 are denoted by $n_1$ and $n_2$ respectively. The notation convention here differs from Ref. \onlinecite{lababidi_counter-propagating_2014} which deals with $\alpha=-1/3$.

\begin{figure*}
\centering
\includegraphics[width=0.9\textwidth]{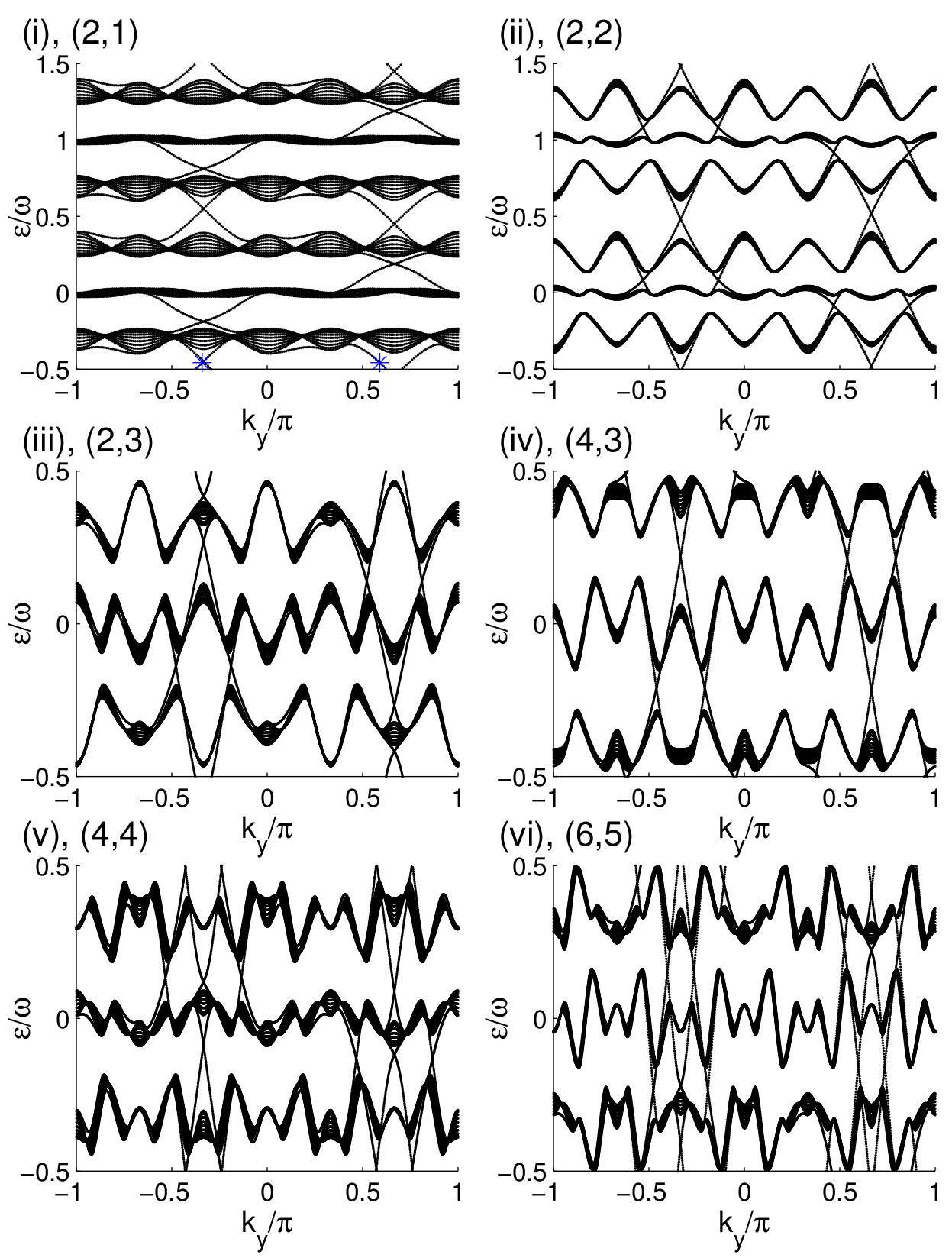}
\caption{
\label{samples}
Representative quasienergy spectra of phase i to vi for the slab geometry. Each phase is labeled by a pair of numbers $(n_1,n_2)$, i.e., at each physical edge there are $n_1$ edge modes, regardless their chirality, in the gap near the QBZ boundary and $n_2$ edge modes in the other two gaps (see main text). From (i) to (vi), the parameters $(j_y,\omega)$ are (1.0, 3.3), (3.3, 6.7), (3.3, 5.0), (4.0, 5.3), (3.3, 3.3), (4.5, 3.3) measured in units of $j_x$. 
}
\end{figure*}

The phases of harmonically driven Hofstadter model can be classified according to the Chern numbers of the bulk bands $(c_1,c_2,c_3)$ and the characteristics of the edge state spectrum, e.g., ($n_1$, $n_2$). The resulting global phase diagram is shown in Fig. \ref{2dpd} on the $(\omega$, $j_y)$ plane for fixed flux $1/3$. For the range of parameters shown, we can see a phase at large driving frequency labeled ``static" and six other distinctive phases labelled by Roman numeral i, ii, up to vi. 
The representative spectra of each phase i to vi, together with the corresponding ($n_1$, $n_2$) values, are shown in Fig. \ref{samples} for a slab with 30 lattice sites in the $x$ direction but infinitely long in the $y$ direction. 
In order to illustrate clearly how the edge states cross the QBZ boundary, we present the spectra of phase i and ii from $-\omega/2$ to $3\omega/2$ in Fig. 2.
The table in Fig. 3 summarizes the characteristics of each phase. Notice that the relation $c_3=c_1$, $c_2=-2c_1$ holds for all seven phases. And the total number of edge states within gap3 is  $n_3=n_2$.

The ``static" phase found here corresponds to phase A of the periodically kicked Hofstadter model studied in Ref. \onlinecite{lababidi_counter-propagating_2014}. Its spectrum (not shown) is topologically identical to that of the static Hofstadter model. For example, there is no edge mode inside the $\pi$-gap. Phase ii corresponds to phase B, while phase iv corresponds to phase D in Ref. \onlinecite{lababidi_counter-propagating_2014}.
Except for the ``static phase", all six other phases, i to vi, have edge modes in the $\pi$ gap, and they always come in pairs, e.g., $n_1=$2, 4 or 6. They are indeed pairs of counter-propagating edge modes with zero net chirality, as discussed in Ref.  \onlinecite{lababidi_counter-propagating_2014}. The edge modes inside the other two gaps have finite net chirality. All these features are in broad agreement with the findings of Ref. \onlinecite{lababidi_counter-propagating_2014} for the periodically kicked Hofstadter model.

We can double check the various relations between the Chern number, the net chirality, and $(n_1, n_2)$ by inspecting the band structure of the truncated Floquet Hamiltonian following the argument of Ref.  \onlinecite{rudner_anomalous_2013}. For the truncated band structure, it is known that the net chirality of edge modes inside any given gap is equal to the sum of the Chern numbers of all the bands below it. 
We have checked numerically that the relation $c_1 + c_2 + c_3 = 0$ holds for all the subbands including those at the edges of the truncation ($m=\pm M$). Accordingly, the net chirality of the edge modes is zero within gap1. And it is equal to $c_3 + c_2$ ($c_3$) for gap2 (gap3).


\begin{figure}
\centering
\includegraphics[width=0.48\textwidth]{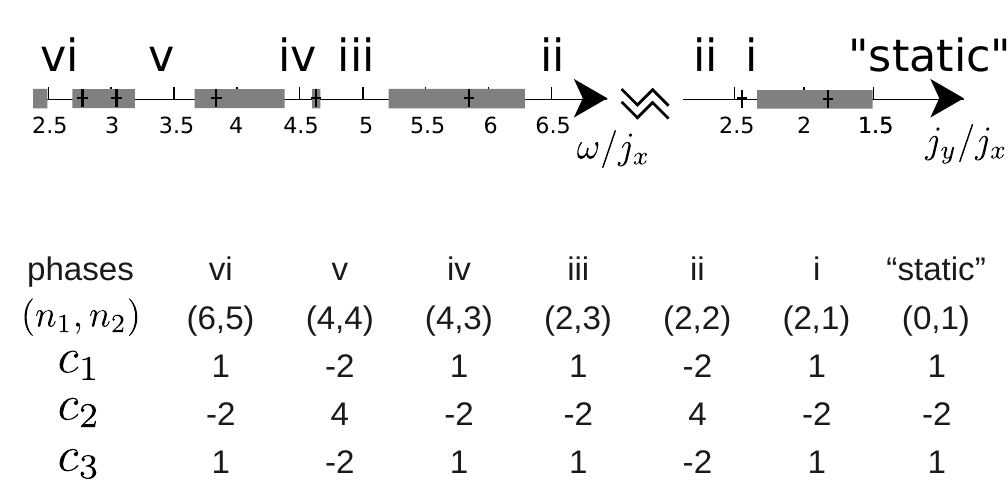}
\caption{
\label{1dpd}
Phases and phase transitions along a particular path in the phase diagram, the dash line in Fig. \ref{2dpd}. White regions are phases with three well defined gap, while the gray ``metallic" regions have at least one of the gaps closed. The table summarizes the Chern numbers, $(c_1,c_2,c_3)$, and the number of edge modes $(n_1, n_2)$ for all the phases i to vi.}
\end{figure}

To examine the successive transitions between these phases, we choose a path on the $(j_y,\omega)$ plane, the dash line in Fig. \ref{2dpd}, that slices through the seven phases, first fixing $j_y=3.3j_x$ and increasing $\omega$, then fixing $\omega=6.7j_x$ and decreasing $j_y$. The resulting one-dimensional phase diagram along this particular path is depicted in Fig. \ref{1dpd}. As in Fig. \ref{2dpd}, white regions are phases with three well defined gaps while the gray regions are ``metallic" where at least one gap closes. The Chern numbers for each phase are calculated from the bulk spectra and listed in Fig. \ref{1dpd} along with the $(n_1, n_2)$ values. Neighboring phases are always separated by a point, labeled by cross (``+") in Fig. \ref{1dpd},  at which two of the bands touch each other at certain values of crystal momentum. We observe that while most of these band touching points are embedded inside a finite metallic region, some phases such as phase i and ii are separated only by a well defined critical point where the gap closes and then reopens immediately. It is important to note that band touching is the necessary but not the sufficient condition for the change in either the Chern number or $(n_1,n_2)$. 
For example, change in $n_1$ is associated with the touching of two bands with the same Chern number, i.e., $c_1$ and $c_3=c_1$. Notice that chiral symmetry demands that the relation $c_3=c_1$ holds even after the band touching and gap reopening. During such a transition, e.g., from the ``static" phase to phase i or from phase iii to phase iv, the Chern numbers do not change. 


To summarize, we have systematically investigated the quasienergy spectrum of the harmonically driven Hofstadter model. It has new phases (such as i, iii, v, and vi) that have not been discussed before. These results suggest that while the phase diagrams differ in details for different driving protocols, the existence of counter-propagating Floquet edge modes seems to be a generic feature of periodically driven quantum Hall systems. 

\section{Wavefunction of the $\pi$-modes}
\label{wf}

\begin{figure}
\includegraphics[width=0.46\textwidth]{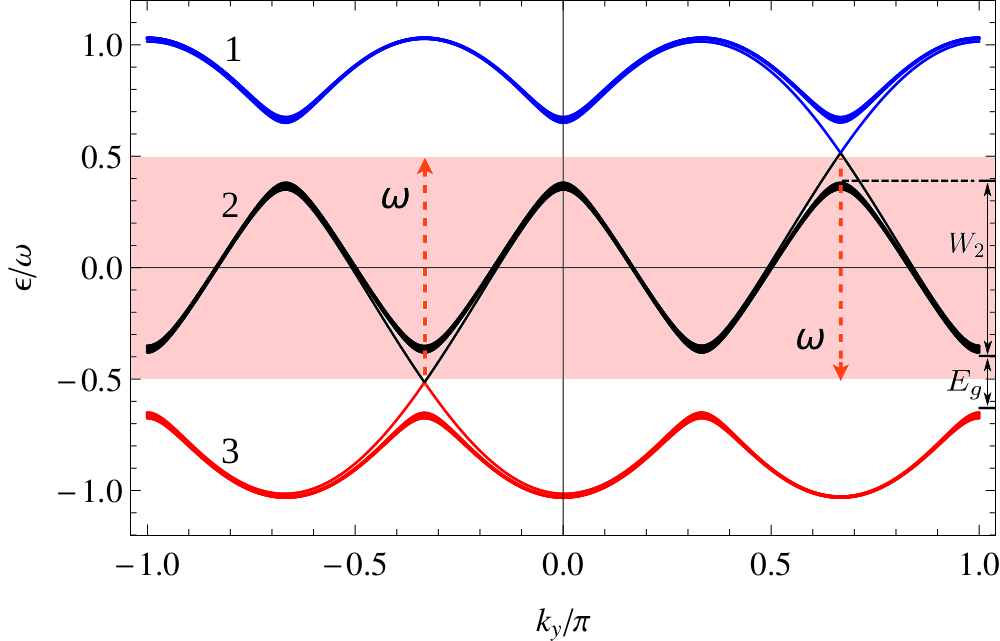}
\includegraphics[width=0.46\textwidth]{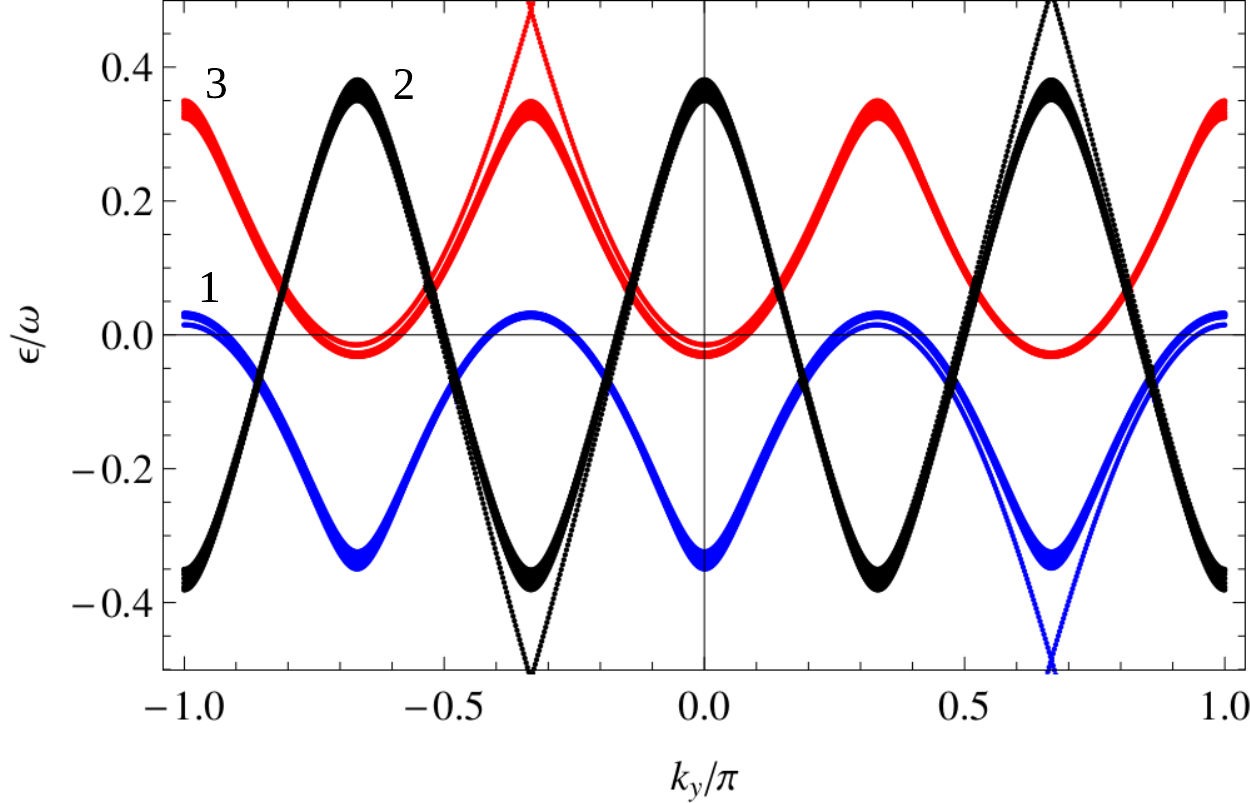}
\includegraphics[width=0.46\textwidth]{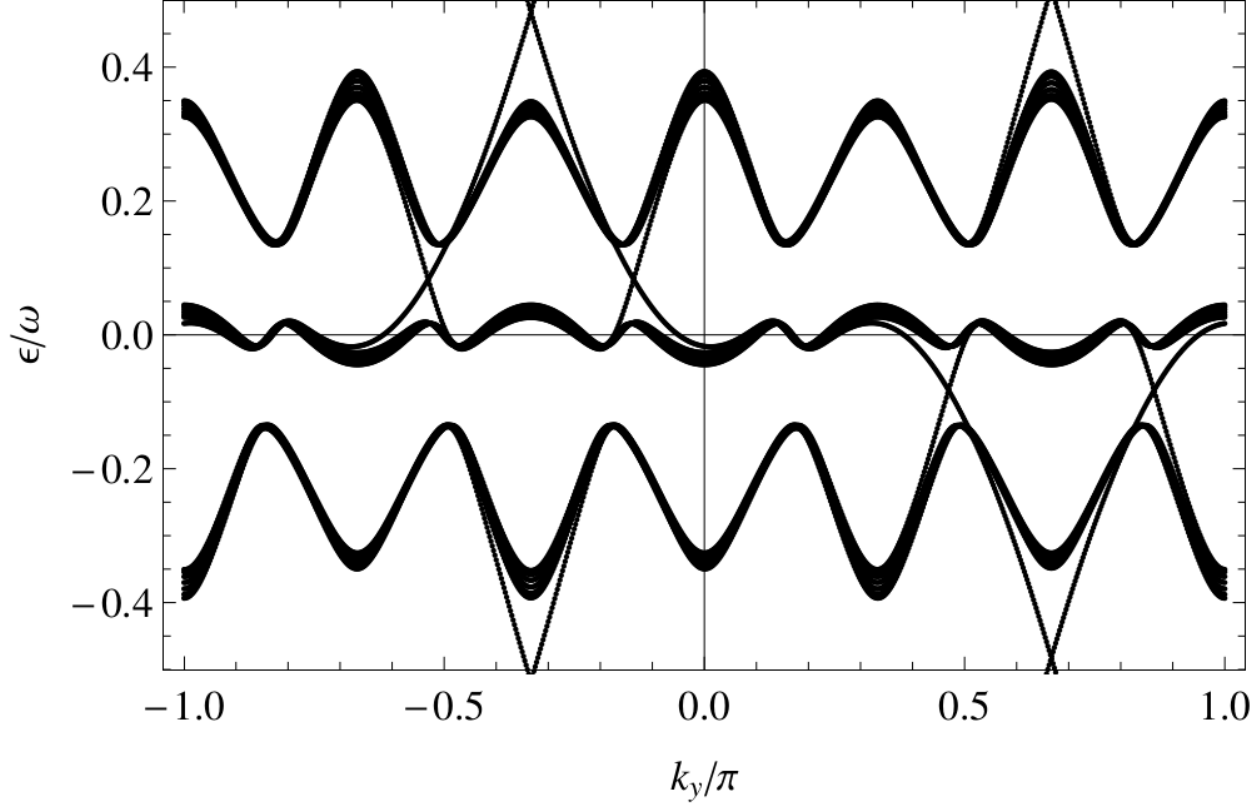}
\caption{
\label{pi_gap}
The evolution of the spectrum for fixed driving frequency as the modulation strength $\gamma$ is increased. Upper panel: the spectrum of the static Hofstadter model, $\gamma=0$. Middle panel: the spectrum of a weakly driven system, $\gamma=0.01$. It nearly coincides with the static spectrum folded into the QBZ (the shaded region). Lower panel: the spectrum of a strongly driven system, $\gamma=1$. $\omega=2j_y=6.7j_x$, $\alpha=1/3$.
 }
\end{figure}

Harmonic driving Eq. \eqref{mod} yields a simple, intuitive interpretation of the counter-propagating edge modes in the limit of weak driving. For simplicity, let us consider the example of phase ii, which has a pair of counter-propagating edge modes inside the $\pi$-gap. To establish its connection with the familiar chiral edge modes of the static Hofstadter model, we imagine gradually increasing the modulation strength $\gamma$ from 0 to 1 for fixed driving frequency $\omega=6.7j_x$ and $j_y=3.3j_x$. The spectrum for $\gamma=0$, $0.01$, and $1$ are shown in the upper, middle and lower panel of Fig. \ref{pi_gap} respectively. 
The quasienergy spectrum for small but nonzero $\gamma$ (weak driving) can be approximated by folding the static spectrum (upper panel) into the QBZ. 
To aid the visualization of the folding, we have labeled the three bands by 1, 2, 3 and indicated the size of the QBZ (the shaded region) in the static energy spectrum. The folded static spectrum nearly coincides with the quasienergy spectrum for $\gamma=0.01$ shown in the middle panel, which has the essential features of the $\pi$-modes. Most importantly, the number of edge modes per physical edge has doubled. This can be seen by focusing on certain quasienergy near QBZ boundary, for example, at $\epsilon=-0.46\omega$. To a good approximation, we can view one of the $\pi$-modes at $\epsilon=-0.46\omega$ as nothing but the edge mode in the static case, i.e., the $m=0$ subband. The other $\pi$-mode can be viewed as deriving from the edge mode originally living in a different gap by absorbing or emitting a quantum of the driving field, a virtual ``photon" of energy $\omega$. Such processes are indicated by the dashed arrows in Fig. \ref{pi_gap}. Symmetries of the Hofstadter model ensures that these two modes have opposite group velocity so that the net chirality is always zero, as already discussed in Ref. \onlinecite{lababidi_counter-propagating_2014}. Note that further increasing the strength of $\gamma$ will hybridize the three bands, leading to additional gap openings and new edge modes away from the QBZ boundary (lower panel for $\gamma=1$). However, the basic features of the $\pi$-modes remain the same. 

The folding construction in the weak driving limit also offers a way to estimate critical driving frequency needed for counter-propagating modes to appear. For example, as long as $\omega$ is larger than the overall band width of the static Hofstadter model, the driven system will remain in the ``static" phase. The $\pi$-modes appear when $\omega$ is within the interval $W_2<\omega<W_2+2E_g$, where $W_2$ is the width of the central band and $E_g$ is the size of the band gap of the static Hofstadter model, see upper panel of Fig. \ref{pi_gap}.

\begin{figure}[t]
\includegraphics[width=0.48\textwidth]{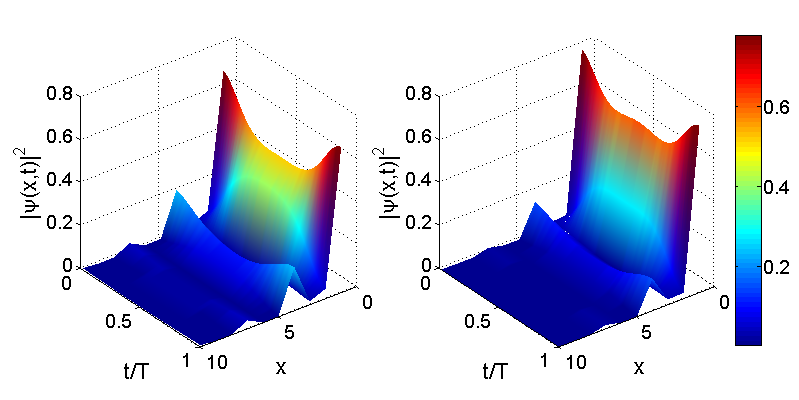}
\caption{
\label{edge-t}
Wavefunction $\psi(x,t)$ of the two counter-propagating $\pi$-modes at the same quasienergy $\epsilon=-0.46\omega$ localized at the same edge $x=1$. They correspond to the star $*$ indicated in phase i of Fig. \ref{samples}(i). The first mode (left panel) is at $k_y=-0.34\pi$ and the second mode (right panel)  is at $0.59\pi$. 
}
\end{figure}

\begin{figure}[h!]
\includegraphics[width=0.48\textwidth]{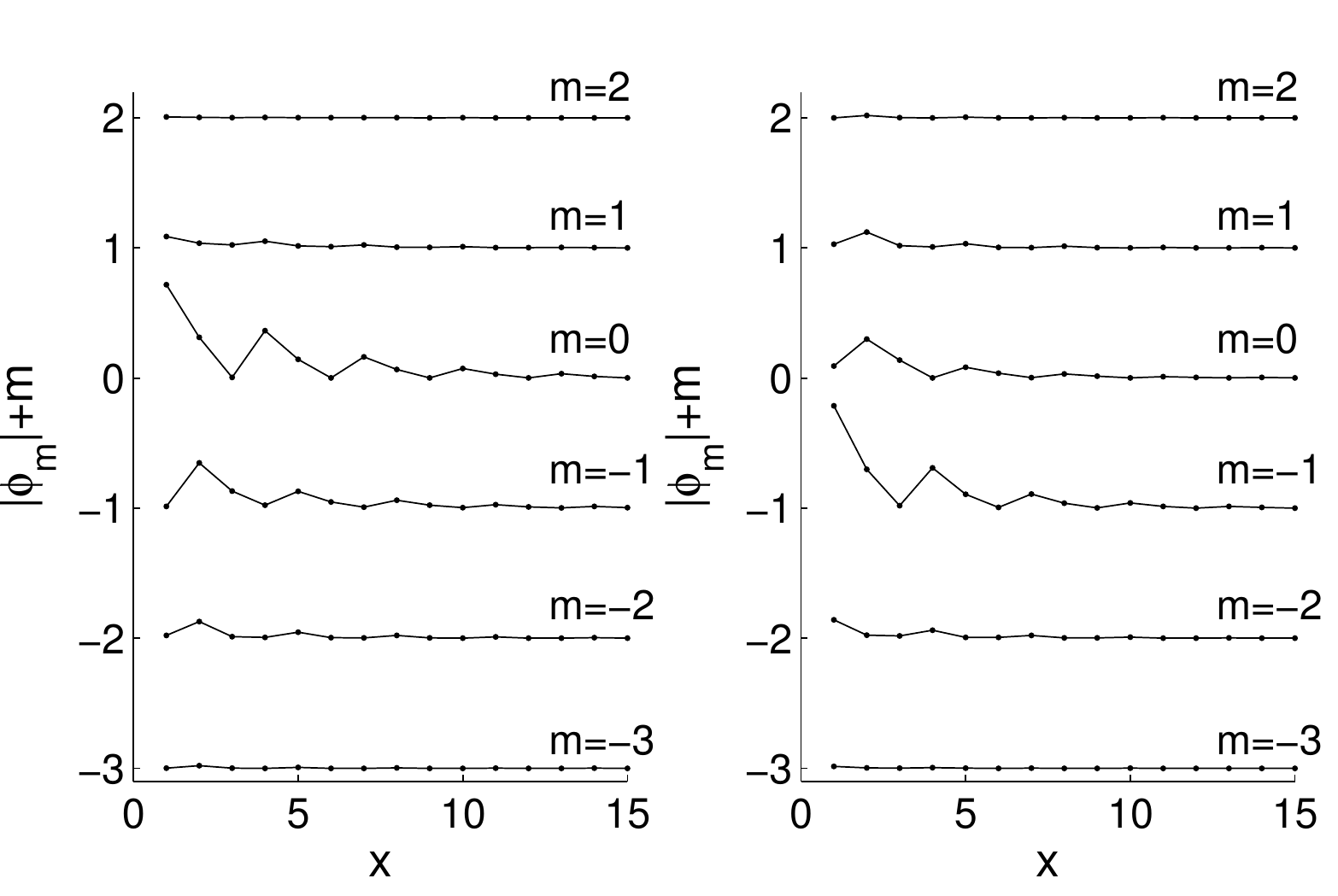}
\caption{
\label{edge-m}
The corresponding frequency components of the edge state wave function shown in Fig. \ref{edge-t}. Different Fourier components, $\phi_m$ for $m=0,\pm1,\pm2...$, are shifted vertically for clarity.
}
\end{figure}

We can put such interpretation of the $\pi$-modes on a firm ground by explicitly analyzing the wave function of the $\pi$-modes outside the perturbative region $\gamma\rightarrow 0$. Fig. \ref{edge-t} shows the wavefunctions $\psi(x,t)$ of the $\pi$-modes localized at edge $x=1$ at $\epsilon=-0.46\omega$ for a strongly driven system, $\gamma=1$, $j_y=j_x$ and $\omega=3.3j_x$ in phase i. The left and right panel correspond to $k_y=-0.34\pi$ and $k_y=0.59\pi$ respectively. These two $k$-points are labeled with $*$ in Fig. \ref{samples}(i). The modulus square of the wavefunction varies with time $t$, attesting the dynamic nature of these Folquet edge modes. We further decompose $\psi(x,t)$ into Fourier series,
\begin{equation}
\label{FT2}
\psi(x,t)=\sum_{m=-\infty}^{\infty}\phi_{m}(x)e^{im\omega t},
\end{equation}
and plot the modulus of the $m$-th Fourier component $\phi_m(x)$ in Fig. \ref{edge-m} for both modes. 
We observe that both $\pi$-modes contain only two appreciable frequency components that differ in $m$ by one. For example, right at the edge $x=1$, $|\phi_m|$ for $m = 1, 0, -1, -2$ are 
\[
0.089,\; 0.717,\; 0.012,\; 0.022
\] 
for the first mode at $k_y=-0.34\pi$ which propagates along $\hat{y}$, and
\[
0.030,\; 0.092,\; 0.787,\; 0.141
\]
for the second mode at $k_y=0.59\pi$ propagating along $-\hat{y}$. We see that the most dominant component of the first mode is $m=0$, while
the most dominant component of the second mode is $m=-1$. That the main contributions to the pair of $\pi$ modes come from two different but neighboring subbands is in broad agreement with our perturbation theory above. Therefore, we can approximately write $\psi_1(x=1,t)\sim (\delta_1 e^{i\omega t}+1)$ for the first mode and $\psi_2(x=1,t)\sim (1+\delta_2 e^{-i\omega t})e^{-i\omega t}$  for the second mode, where only the leading terms are retained and $\delta_1,\delta_2\ll 1$ are small numbers. In both cases, $|\psi_{1,2}(x=1,t)|^2\sim 1+2\delta_{1,2}\cos(\omega t)$, in qualitative agreement with the time dependence of $\psi$ shown in Fig. \ref{edge-t}. 

This crude caricature of the $\pi$-modes not only provides a simple, intuitive picture to understand their frequency components and time dependence, but also indicates their robustness against weak static perturbation $V_0$. $V_0$ only couples degrees of freedom within the same subband. To the leading order (neglecting the subdominant frequency components), the matrix element of $V_0$ between $\psi_1$ and $\psi_2$ is zero. Hybridization of the two $\pi$-modes requires a potential $V(t)$ that couples the $m=0$ subband and the $m=-1$ subband, i.e., $V(t)$ has to have a characteristic frequency $\omega$. In the next section we numerically check the robustness of the counter-propagating edges modes against a variety of static perturbations.

\section{Robustness of the $\pi$-modes}

\begin{figure}
\centering
\includegraphics[width=0.48\textwidth]{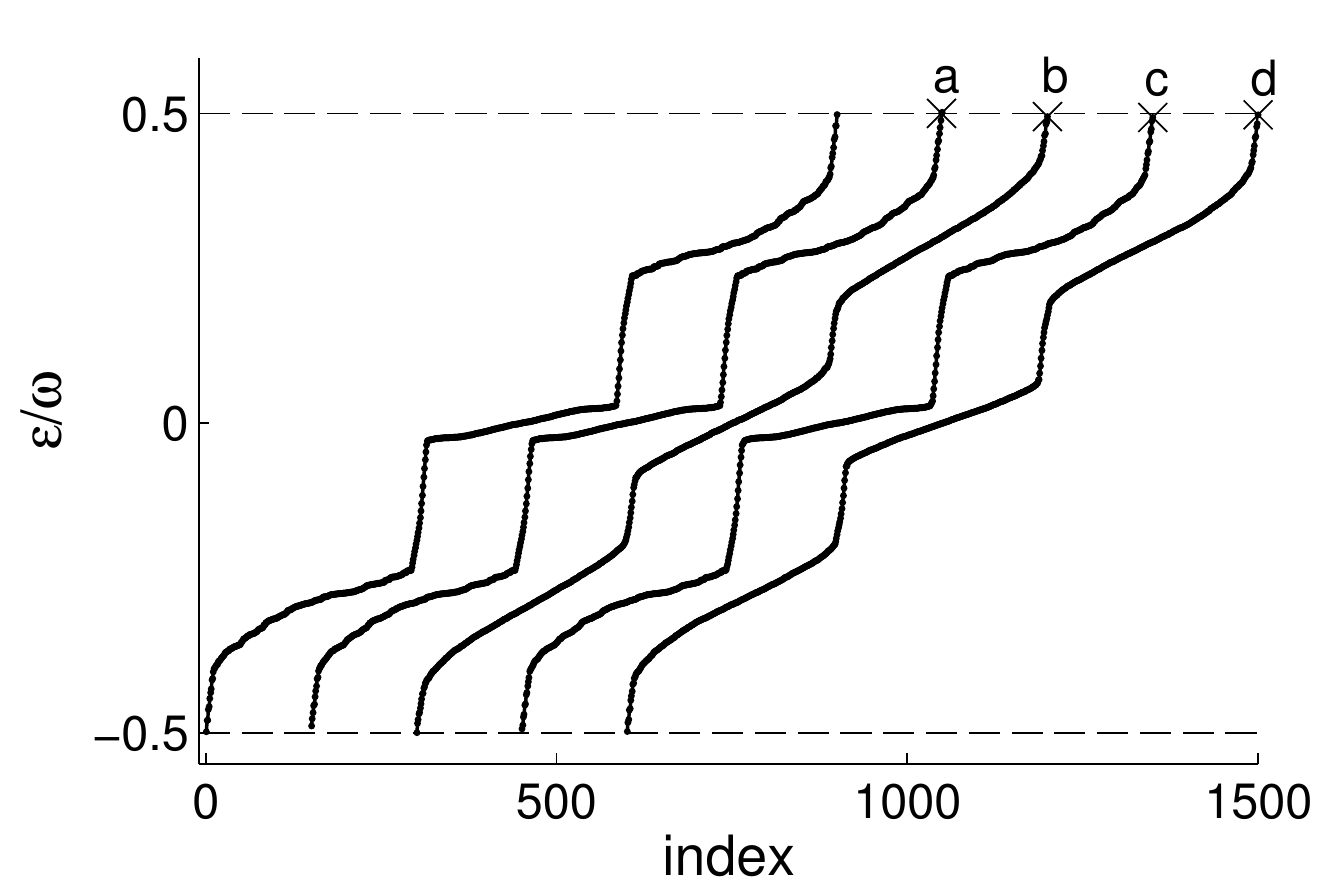}
\caption{
\label{eng_imp}
The quasienergy eigenvalues of harmonically driven Hofstadter model on a finite 30$\times$30 lattice for $j_y=j_x, \omega=3.3j_x$ (phase i), $\alpha=1/3$ and $\gamma=1$.
The horizontal axis is the index of the eigenvalues. The leftmost is the spectrum without perturbation. The remaining spectra, a to d, are 
for different perturbations (see main text).
They are shifted horizontally from each other for clarity. Static onsite or bond disorder does not open a gap in the spectrum, e.g., at the QBZ boundary (dash line) $\epsilon=\pm\omega/2$.
}
\end{figure}

\begin{figure*}
\centering
\includegraphics[width=0.8\textwidth]{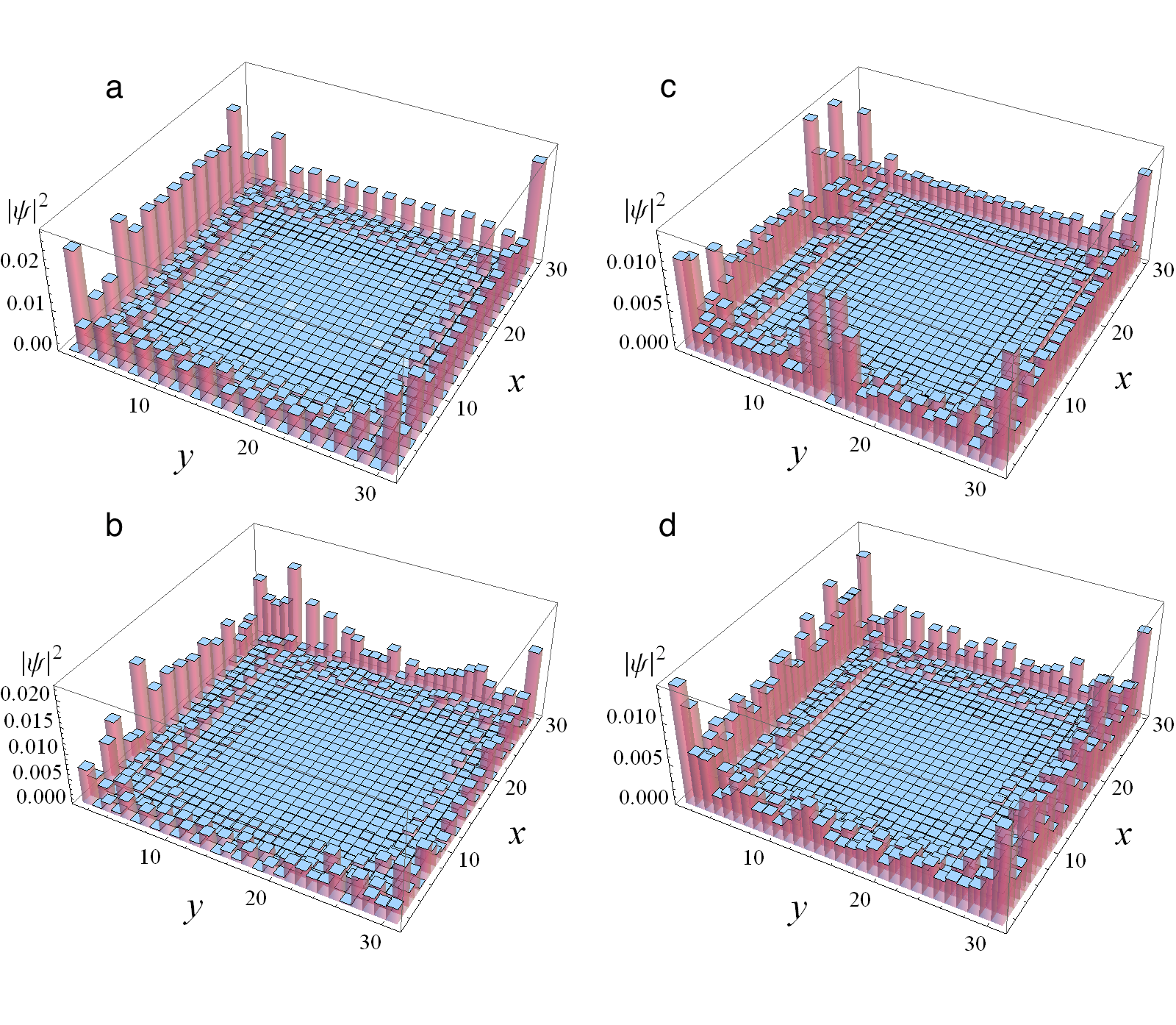}
\caption{
\label{imp}
The wave function $\psi(x,y,t=0)$ of the $\pi$-modes in the presence of disorder. The parameters used for panel a, b, c, d are shown as $\times$ in Fig. \ref{eng_imp}. They all corresponds to quasienergy very close to $\epsilon=\omega/2$. These plots offer direct evidence for the robustness of the $\pi$-modes against static disorder.}
\end{figure*}

We have numerically computed the quasienergy spectrum of a finite lattice of Hofstadter model under harmonic driving for a variety of static perturbations. These include (a) an exceedingly large onsite potential ($10^3j_x$) at two sites on the edge $(x,y)=(22,1)$ and $(1,15)$; (b) random onsite disorder $\delta\mu\in[-0.5j_x, 0.5j_x]$; (c) zero hopping to and from the same two sites on the edge as in (a); and (d) random hopping disorder $\delta j_{x,y}\in [-0.2j_x,0.2j_x]$. The results are shown in Fig. \ref{eng_imp} and compared to the original (the leftmost) spectrum  before any perturbation being introduced. Despite the rather strong perturbations, there is no gap opening in the spectrum. The counter-propagating edge modes are robust against local defects or random distribution in either onsite potential or hopping between the neighboring sites. To prove this directly, we plot in Fig. \ref{imp} the real space wavefunction $\psi(x,y,t=0)$ corresponding to quasienergy eigenvalue $\epsilon\simeq \omega/2$, i.e., the points labeled with $\times$ in Fig. \ref{eng_imp}. They clearly are edge states. Compared to the unperturbed edge states, disorder distorts the wavefunction but does not destroy the $\pi$-modes. 

\section{Discussion}
Our analysis above on the harmonically driven Hofstadter model provides additional evidence that counter-propagating edge modes are generic features of periodically driven integer quantum Hall systems. They survive the introduction of static disorder and appear rather robust. 
The fact that they arise for very different driving protocols is another manifestation of their robustness: variations in the driving details will not destroy them. We can go one step further using perturbative arguments in Section \ref{wf}. A static quantum Hall insulator is known to be robust against any weak perturbations $V_0$, such as disorder, provided that the $U(1)$ gauge symmetry associated with the conservation of fermion particle number is not broken. In particular, its edge states and band gaps will survive these perturbations. Now turn on a weak harmonic driving. According to the folding construction which holds even in the presence of $V_0$, pairs of counter-propagating edge modes naturally arise when $\omega$ falls into the interval between $W_2$ and $W_2+2E_g$ as discussed in Section \ref{wf} despite the presence of $V_0$. 
In this sense, the robustness of the $\pi$-modes is rooted in the robustness of the edge state in the integer quantum Hall effect.
While this argument is limited to the weak driving limit, our numerical results suggest that the same conclusion hold for larger modulation strength $\gamma$.

This is in sharp contrast with the conventional wisdom regarding the edge state of static two-dimensional systems which states that a pair of counter-propagating edge modes living on the same edge tend to be unstable without the protection of time-reversal symmetry. The main point coming out of this and our earlier work \cite{lababidi_counter-propagating_2014} can be summarized as the following sentence. There exist interesting, robust edge state phenomena in periodically driven two-dimensional systems, even when the winding number (the net chirality) is zero. Our conjecture is that there exists another topological invariant that can predict the number of edge mode for each chirality. Explicit construction of such invariant warrants a separate work.

\begin{acknowledgements}
This work is supported by AFOSR FA9550-12-1-0079 (ZZ and EZ) and NSF PHY-1205504 (EZ). We thank Mahmoud Lababidi and Vincent Liu for helpful discussions.
\end{acknowledgements}

\bibliography{pdriving1}

\end{document}